\documentclass[twocolumn,superscriptaddress,showpacs,showkeys,preprintnumbers,aps]{revtex4}

\usepackage{amssymb}
\usepackage{amsmath}
\usepackage{graphicx}
\usepackage{float}
\usepackage{color}

\begin{document}

\title{\boldmath Chemical disorder as engineering tool for
  spin-polarization\\ in Mn$_3$Ga-based Heusler systems}

\author{S.~Chadov}
\affiliation{Max-Planck-Institut f\"ur Chemische Physik fester Stoffe,
  01187  Dresden, Germany}
\author{S.~W.~D'Souza}
\affiliation{Max-Planck-Institut f\"ur Chemische Physik fester Stoffe,
  01187  Dresden, Germany}
\author{L.~Wollmann}
\affiliation{Max-Planck-Institut f\"ur Chemische Physik fester Stoffe,
  01187  Dresden, Germany}
\author{J.~Kiss}
\affiliation{Max-Planck-Institut f\"ur Chemische Physik fester Stoffe,  
01187  Dresden, Germany}
\author{G.~H.~Fecher}
\affiliation{Max-Planck-Institut f\"ur Chemische Physik fester Stoffe,
  01187  Dresden, Germany}
\author{C.~Felser}
\affiliation{Max-Planck-Institut f\"ur Chemische Physik fester Stoffe,
  01187  Dresden, Germany}

\email{stanislav.chadov@cpfs.mpg.de}


\begin{abstract}
Our study highlights spin-polarization mechanisms in metals, by focusing
on the mobilities of conducting electrons with different
spins instead of their quantities. Here, we engineer electron mobility by applying chemical disorder induced by 
non-stoichiometric variations. As a practical example, we discuss 
the scheme that establishes such variations in tetragonal Mn$_3$Ga
Heusler material. We justify this approach using  first-principles calculations of the spin-projected
conductivity components based on the Kubo-Greenwood formalism. It
follows that, in majority of the cases, even a small substitution of some other
transition element instead of Mn may lead to a substantial increase in spin-polarization along the tetragonal axis.
\end{abstract}
\maketitle
\section{Introduction}

Improved understanding of the influence of disorder in solids yields
potential new approaches to the state-of-the-art design of multicomponent materials. Despite
the popular view that  disorder should be avoided by any means, one can 
find its constructive influence on physical properties in various applied
fields. Examples include  the disorder-induced mechanical work hardening
of materials~\cite{SH04}, ``softer'' examples such as the
efficient blocking of sound waves in liquid foams~\cite{PDL14}, and 
the Anderson localization of light in disordered photonic crystals~\cite{STS+10}. By carefully reviewing the literature,
one finds many more constructive examples. Constructive disorder simply prevents the propagation
 of certain quasiparticles. Indeed, in the first example,  mechanical
 dislocations are  blocked by other type of random defects; in the second, ultrasonic
 phonons are blocked by random foam bubbles; and in the third,
 electromagnetic waves are blocked
 by the breaking of the translational symmetry in the photonic crystal. Of course,
 similar phenomena also accompany electronic propagation in metals, where
 the breaking of translational symmetry (static or dynamic) leads to a nonzero
 resistivity. Here, one of the most dramatic examples is 
 the Anderson localization~\cite{And58,Tho74}, which completely blocks the propagation
 of an electron despite the absence of a semiconducting gap in the
 density of states. This can have constructive implications,
 for example, it has recently been proposed that Anderson
 localization induced in topologically non-trivial systems, such
 as HgTe-type semimetals, should result in a new class of topological
 insulator~\cite{LCJS09,GWA+09,JWSX09}. On the other hand, in ``typical''
 metals (i.e.,  systems with a conducting electron density of ${n_{\rm F}\gtrsim5\times10^{22}}$~cm$^{-3}$), the Anderson-Mott criterion~\cite{Mot36} (${a_{\rm
B}n^{1/3}_{\rm F}<0.25}$, where $a_{\rm B}$ is the Bohr radius) cannot be
 fulfilled,  so an unlimited increase in disorder leads to a saturation of metallic resistivity
 which, in practice, is  restricted by the empirical Mooij limit, $\rho_{\rm max}\lesssim300~\mu\Omega$cm~\cite{Moo73}.

Here, we would like to demonstrate another interesting effect 
that can be induced by disorder in metallic systems, the
so-called {\it spin-selective} electron localization. Specifically, we will
justify the  possibility of creating such a type of disorder, which 
noticeably localizes the conducting electrons of one spin but almost
negligibly affects the conduction of the other spin. It is rather
clear that, since the spin subbands in a magnetic metal are different,
their conductivities also differ (i.e. ${\sigma^{\uparrow}\neq\sigma^{\downarrow}}$). The extreme case, 
 which is especially interesting in terms of spintronics, is the half-metallicity
(i.e. ${\sigma^{\uparrow}>0}$, ${\sigma^{\downarrow}=0}$ or vice
versa) characterized by the highest possible amplitude of the spin-polarization,
${P=\frac{\sigma^{\uparrow}-\sigma^{\downarrow}}{\sigma^{\uparrow}+\sigma^{\downarrow}}=\pm1}$. It
can be realized in the special class of materials
known as half-metals~\cite{KWS83,GMEB83}, which possess a semiconducting
band gap in one spin channel only (i.e. ${n_{\rm F}^{\downarrow}=0}$, ${n_{\rm F}^{\uparrow}>0}$ or vice
versa). However, as one can see in the simple Drude picture,
${\sigma\sim n_{\rm F}l}$. Thus, the conductivity also scales with an
electronic mean free path $l$, providing the potential for ${P=\frac{n_{\rm F}^{\uparrow}l^{\uparrow}-n_{\rm F}^{\downarrow}l^{\downarrow}}{n_{\rm F}^{\uparrow}l^{\uparrow}+n_{\rm F}^{\downarrow}l^{\downarrow}}}$,
to be adjusted by  manipulation of the electron mobilities, $l^{\uparrow(\downarrow)}$, in
different spin channels, rather than adjusting the $n_{\rm
  F}^{\uparrow(\downarrow)}$ only. 

A large number of mechanisms exist that favor the disorder 
in solids provided by a diverse manifold
of the degrees of freedom, whether thermal (such as phonons, magnons,  polarons) or
fully intrinsic (such as through geometrical frustration, stoichiometric
variations and electron interaction). In order to engineer such
mechanisms efficiently, one must understand their impact
on the electronic structure. At present, certain connections can be established
using special mean-field theories (e.g. coherent potential
approximation (CPA)~\cite{Sov67,But85} or dynamical mean-field theory (DMFT)~\cite{MV89,KSH+06}), statistical methods (e.g. Monte-Carlo-based simulations) or
combined approaches. 

Here, we will restrict our engineering to a very fundamental level by improving the ground-state electronic transport
characteristics through a particular type of chemical disorder, as it is 
one of the most common phenomena in polyatomic solid compositions.
The simplest way to introduce chemical disorder is through  variation
of the stoichiometry. Here, a convenient test environment is
provided by the Heusler family of materials, which typically have a ternary
composition: two different transition metals (TM) and one main-group
element (MG). The majority of these substances crystallize in the fcc-based cubic structure
(centrosymmetric $\textit{Fm-3m}$ or noncentrosymmetric $\textit{F-43m}$) and obey the same chemical
ordering rules~\cite{Web69}. By substituting one TM for another TM or one MG by another MG, the properties of
Heusler materials can be varied  widely without affecting their structure,
 from nonmagnetic/semiconducting to magnetic/metallic. In particular,
the latter class includes the majority of the known half-metals as given
in Refs.~\cite{KWS83,GMEB83}. A nonstoichiometric substitution will lead
to a random occupation of the corresponding Wyckoff sites, by automatically
breaking the translational invariance. This random site occupation by
two or more elements with localized electronic subbands (e.g. $3d$ states)
centered at different energies will disturb the coherent scattering of the
delocalized (mobile) electrons at corresponding energies and result in
their partial localization. Thus, if the nonstoichiometric substitution
induces these random fluctuations within the energy window containing
the Fermi level ($E_{\rm F}$), the resistivity will increase. The
concept is therefore clear; we must create  such an
energy window at $E_{\rm F}$ in one spin channel, and simultaneously shift it away from $E_{\rm F}$  in
the other channel. Such a situation can be maintained, obviously, only in magnetic systems.

\section{Random fluctuation design:\ \  Mn$_3$Ga and its derivatives}
\label{sec:Random_fluctuations_design}

\begin{figure}
\centering
\includegraphics[clip,width=1.0\linewidth]{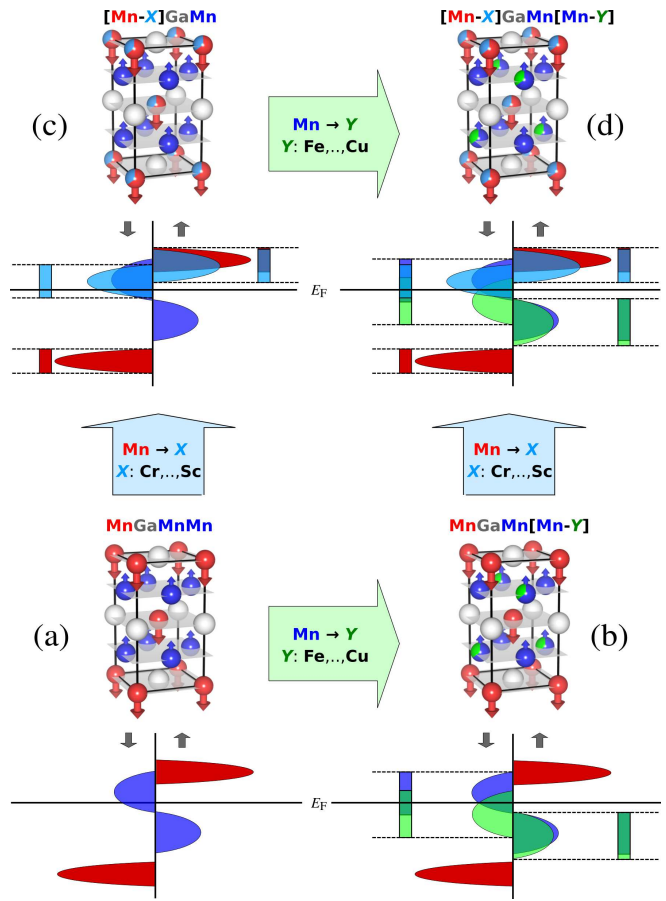}
\caption{\label{FIG:scheme} (a)~Mn$_3$Ga unit cell: Mn atoms
  in the $2a$ (red) and $2c/2d=4c$ (blue) Wyckoff sites; Ga atoms are in $2b$ (gray); arrows indicate atomic magnetic
  moments. According to the sequence of Wyckoff positions,
  $2a\,2b\,2c\,2d$, we label this compound as MnGaMnMn.
(b)~Substituting a late TM in place of Mn   ($Y$, light-green) results in a MnGaMn[Mn-$Y$]
  alloy  with  Mn-$Y$ disorder on $2d$ sites. (c)~Replacing Mn with an
  early TM  ($X$, light-blue) results in a [Mn-$X$]GaMnMn alloy with Mn-$X$
  disorder on $2a$. (d)~Simultaneous combination of (b) and (c):
  [Mn-$X$]GaMn[Mn-$Y$].  The corresponding spin-projected and
  atomic-resolved $3d$-DOS scheme for each prototype system
  (left-oriented peaks are spin-down, right-oriented are
  spin-up, as indicated by gray arrows). The atom-projected DOS contributions
  obey the same color code as that used for the atoms; vertical colored bars indicate the energy windows in
  which the electronic levels randomly fluctuate.}
\end{figure}

We select Mn$_3$Ga Heusler as a suitable object, since it possesses strong
local moments and is not a half-metal. This is because of its tetragonal distortion which reduces
its point symmetry to $\textit{I4/mmm}$. Chemically speaking it is a
binary, however, it contains two types of Mn in the $2a$ and $4c$ Wyckoff positions (marked
red and blue, respectively, in  Fig.~\ref{FIG:scheme}\,a) 
which differ by their nearest neighbor environment. Mn$(2a)$ has the largest local
moment (according to different measurements and first-principles
estimations  $\approx$\,3\,-\,4~$\mu_{\rm B}$) coupled antiparallel to Mn$(4c)$,
which has a smaller moment of $\approx$\,2\,-\,3~$\mu_{\rm B}$. Since
the $4c$ class
contains two Mn atoms (in $2c$ and $2d$, which are equivalent), the
total magnetization, $M$, is oriented along the
$4c$ local moments. ${M\approx1.7~\mu_{\rm B}}$/f.u. (f.u.=formula unit) according
to the first-principles calculation~(see, e.g.,~\cite{WBF+08}), and ${M\approx1.1~\mu_{\rm B}}$/f.u. 
according to  experiment~\cite{AWFF11,WCG+12,RBB+13} (note that the saturation was not achieved).

Let us examine the energetic structures of our future scattering centers (localized 3$d$
states of Mn). They are shown schematically in
Fig.~\ref{FIG:scheme}\,a. Mn($2a$), which exhibits the strongest magnetic
moment, is close to the half-filled state. Its $3d$ shell contains five spin-down
electrons, whereas the spin-up states are empty (ideally this should
result in 5~$\mu_{\rm B}$, however, this can differ in calculations,
since the amplitude of the local moment depends on the position at which
the ``border'' between the atoms is set). On the other hand, Mn$(4c)$
exhibits a smaller moment and  weaker exchange split; this simply
means that, whereas the spin-up band is fully filled, the spin-down band is
not fully empty, i.e., it contains $E_{\rm F}$. Such a ``half-metallic''
structure of the localized electronic subsystem (the delocalized $s$ and
$p$ conducting electrons, which are not shown in the scheme, do not have a band gap
in any of the spin channels)  fits well into the framework of the Anderson
impurity model~\cite{And61}, which explains the spin dependence of
the conducting electron scattering on a given magnetic impurity. In the
second-order perturbation theory, the scattering process involves an intermediate
state in which a conducting electron occupies an impurity level. At $E_{\rm F}$, this is a partially filled
spin-down $3d$ subband of Mn$(4c)$. According to the Pauli exclusion
principle, intermediate states in which the impurity level is occupied by two
electrons with the same spin orientation are forbidden. Thus, the
spin-down conducting electrons will be repelled from the Mn$(4c)$ spin-down
subband more strongly, which results in a relative increase in the spin-down resistivity component,
$\rho^{\,\downarrow}$. The only ingredient which is still missing is
the chemical disorder on the $4c$ sites, which leads to a random
fluctuation of the spin-down localized  $d$-electron subband by causing
it to become an efficient scattering center.

This type of disorder can be introduced by substituting some other TM in
place of Mn. One of the general rules, that holds quite unambiguously for Heusler alloys concerns their chemical ordering~(see
e.g. \cite{Web69}). Specifically, the {\it earliest} TM (which is located closer to the left side of the periodic table,
i.e., which belongs to the earliest group) shares the same atomic layer
as the MG element (e.g., if Ga occupies 
$2b$, the earliest TM occupies the  $2a$ Wyckoff site). Thus, for Mn-$Y$
substitution with the {\it later} TMs, $Y$ = Fe, Co, Ni, Cu,  the earliest TM is Mn, which
therefore remains in $2a$. For this  reason,  $Y$ will randomly occupy 
 the $4c$ Wyckoff sites. As follows from the present calculations 
(in the case of $Y$ = Fe, Co, Ni and also experimentally~\cite{AWFF11,WCG+12}), 
such substitution preferably occurs  on one of the two $4c$ sites ($2c$
or $2d$; here, let us choose  $2d$) by rendering them nonequivalent (see
Fig.~\ref{FIG:scheme}\,b). According to the sequence $2a\,2b\,2c\,2d$,
we label the resulting compound  MnGaMn[Mn$_{1-y}Y_y$]. Statistically, its point symmetry 
reduces from $\textit{I4/mmm}$ to $\textit{I-4m2}$ (no inversion). Due to
similarity with Mn$(2c/2d)$ the magnetic moments of $Y(2d)$ are coupled
negatively to Mn$(2a)$, i.e., they are oriented up.  Since $Y(2d)$
has a smaller magnetic moment compared to Mn$(2c/2d)$ (because of its more complete
$d$-shell), its partially occupied spin-down subband is downshifted
 energetically with respect to Mn($2c/2d$), whereas its fully occupied
spin-up subband is centered below  $E_{\rm F}$, similar to the spin-up
subband of Mn$(2c/2d)$. It is clear that such substitution yields the
spin-down random fluctuation window including $E_{\rm F}$, whereas the
spin-up window is situated below $E_{\rm F}$  (see Fig.~\ref{FIG:scheme}\,b). 

A similar effect can be maintained by replacing Mn with earlier TMs: $X$
= Cr, V, Ti and Sc. According to the aforementioned chemical ordering
rules, the earlier TMs  occupy the $2a$ Wyckoff
site, which is represented in formal notation as [Mn$_{1-x}X_x$]GaMnMn (see
Fig.~\ref{FIG:scheme}\,c). This does not change the $\textit{I4/mmm}$
point symmetry. Because of the similarity with Mn$(2a)$, the
magnetic moments of $X(2a)$ are also coupled negatively to Mn$(4c)$, i.e. are
down-oriented. Since the $d$-shell of the $X$ element is less
than half-filled, its spin-up subband is fully empty
(it is situated above $E_{\rm F}$), whereas the spin-down subband is partially
filled (contains $E_{\rm F}$). This again provides two randomly
fluctuating energy windows: one including $E_{\rm F}$, in the spin-down 
channel, and the other one, above $E_{\rm F}$ in the spin-up channel. An
additional energy regime of random fluctuations which occurs in the
spin-down channel relatively far below $E_{\rm F}$ exists, which is caused by
the deepest fluctuating spin-down subband of Mn($2a$), as shown in Fig.~\ref{FIG:scheme}\,c.
Since both substitutions [Mn$_{1-x}X_x$]GaMnMn and MnGaMn[Mn$_{1-y}Y_y$]
are independent, they can be simultaneously combined into
[Mn$_{1-x}X_x$]GaMn[Mn$_{1-y}Y_y$], as shown in
Fig.~\ref{FIG:scheme}\,d. The resulting compound statistically corresponds to the
lowest $\textit{I-4m2}$ symmetry, and its fluctuation spectrum represents
 a superposition of the energy windows in cases (b) and (c).

\section{First-principles  justification}

\subsection{Visual analysis of the Bloch spectral\ \ \ \ \ \  function (BSF)}

We assume, that the substitution rate, $x$  or $y$, must be
sufficiently small everywhere to prevent noticeable structural or
electronic changes. Experimentally, in the case of $Y$ = Co, the critical concentration
(when the Mn$_{3-y}Y_y$Ga composition reaches into the cubic phase) is
${y\approx0.5}$; in the case of $Y$ = Fe or Ni, it is at least
${y=1}$~\cite{WCG+12}. To date, cases of Mn-$X$ substitution have not been reported
apart from $X$ = V, known to be cubic for ${x=1}$~\cite{KKM+08}. For this
reason, among all the cases considered here, the practically interesting
examples are those in which the substitution rate does not exceed ${10-20}$\,\%
(${0<x,y\lesssim0.2}$). Nevertheless, since it is interesting  to also
track the spin-polarization at the disorder rate  maximum, we will study Mn$_{3-y}Y_y$Ga and Mn$_{3-x}X_x$Ga within a wide
range (${0\le x,y\le0.5}$). To examine the effects  of the combined  substitution
(in Mn$_{3-x-y}X_xY_y$Ga) we will take ${x=y}$ but for consistency we will
ensure that the sum of these rates does not exceed the maximal rate (${x+y\le0.5}$). All relevant computational
details, including the definition of the Bloch spectral function (BSF)
used in the following, are listed in Sec.~\ref{sec:appendix}.

\begin{figure}
\centering
\includegraphics[clip,width=1.0\linewidth]{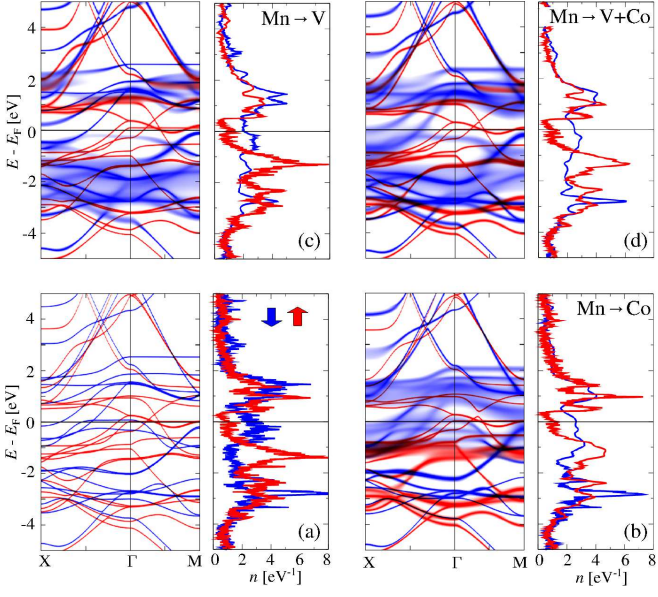}
\caption{\label{FIG:BSF-DOS} Spin-projected BSF (red and blue indicate
  spin-up and spin-down, respectively)  along the X-$\Gamma$-M path and the spin-projected total DOS
  computed for (a)~$\text{MnGaMnMn}$, (b)~MnGaMn[Mn$_{0.5}$Co$_{0.5}$],
  (c)~[Mn$_{0.5}$V$_{0.5}$]GaMnMn, and
  (d)~[Mn$_{0.25}$V$_{0.25}$]GaMn[Mn$_{0.25}$Co$_{0.25}$], corresponding to
  the schemes in Figures\,\ref{FIG:scheme}\,a, \ref{FIG:scheme}\,b, \ref{FIG:scheme}\,c, and \ref{FIG:scheme}\,d, respectively.}
\end{figure}

By comparing the calculated spin-projected BSFs (red and blue indicate
spin-up and spin-down, respectively) of Mn$_3$Ga (Fig.~\ref{FIG:BSF-DOS}\,a) 
with Mn$_{2.5}$Co$_{0.5}$Ga (Fig.~\ref{FIG:BSF-DOS}\,b) and
Mn$_{2.5}$V$_{0.5}$Ga (Fig.~\ref{FIG:BSF-DOS}\,c), one can identify all
the fluctuation regimes (visualized as broadened regions) schematically presented in
Fig.~\ref{FIG:scheme}. In the case of Mn-Co substitution, the random fluctuations
in the spin-down channel span a wide energy window ${\approx E_{\rm F}\pm2}$~eV, whereas no broadening is observed for the
spin-up states crossing $E_{\rm F}$, as the spin-up fluctuations
begin only at 1~eV below $E_{\rm F}$. In the case of Mn-V, the spin-down
fluctuation region is even wider: it roughly spans the $-3.5$ to 2~eV
range as a fluctuating superposition of the lower and higher  bands of Mn($2a$) and
V($2a$), respectively. Again, almost no broadening of
the spin-up states is observed at $E_{\rm F}$, as the spin-up fluctuation window is now
shifted above $E_{\rm F}$ (from $0.5$ to 2~eV). In the combined 
Mn$_{2.5}$V$_{0.25}$Co$_{0.25}$Ga case (Fig.~\ref{FIG:BSF-DOS}\,d) the fluctuation regime
clearly represents a superposition of the fluctuation energy windows in Figs.~\ref{FIG:BSF-DOS}\,c and \ref{FIG:BSF-DOS}\,d.

Before we examine to the quantitative estimates, it is instructive to note the difference in informational content
provided by the BSF and DOS. In all cases, the spin-resolved DOS 
indicates that at $E_{\rm F}$,  ${n^{\uparrow}_{\rm F}<n^{\downarrow}_{\rm
    F}}$, which may lead us to naively assume of a negative
spin-polarization, i.e., ${P\sim n^{\uparrow}_{\rm F}-n^{\downarrow}_{\rm
    F}<0}$. Such an estimate is used
quite often, even today. In certain cases, this estimate can be improved
upon if, instead of the total DOS, only its $s$- and
$p$-electron projections are considered, but even such improvement can
be efficient only when the electron mobility values in both spin channels are
close. In contrast, we see from the BSFs  that the electron momentum uncertainties
(${\Delta k\sim l^{-1}}$) at $E_{\rm F}$ produced by disorder are also
very different for the two spins (${\Delta k^{\uparrow}<\Delta
  k^{\,\downarrow}}$) and suggest the opposite conclusion, i.e., a
positive spin-polarization, ${P\sim l^{\uparrow}-l^{\downarrow}>0}$. 
Obviously, in such situations, a final conclusion can be made only if it is
based on approaches adequately accounting for both factors. For this
reason,  we compute the spin-projected resistivities as
 functions of $x$ and $y$ in Mn$_{3-x}X_x$Ga, Mn$_{3-y}Y_y$Ga and
Mn$_{3-x-y}X_xY_y$Ga alloys in the following, using the Kubo-Greenwood linear response
formalism~\cite{Kub57,Gre58} and applying the relativistic spin-projection scheme~\cite{LKE10}.

\subsection{Quantitative analysis of spin-polarization}

\begin{figure}
\centering
\includegraphics[clip,width=1.0\linewidth]{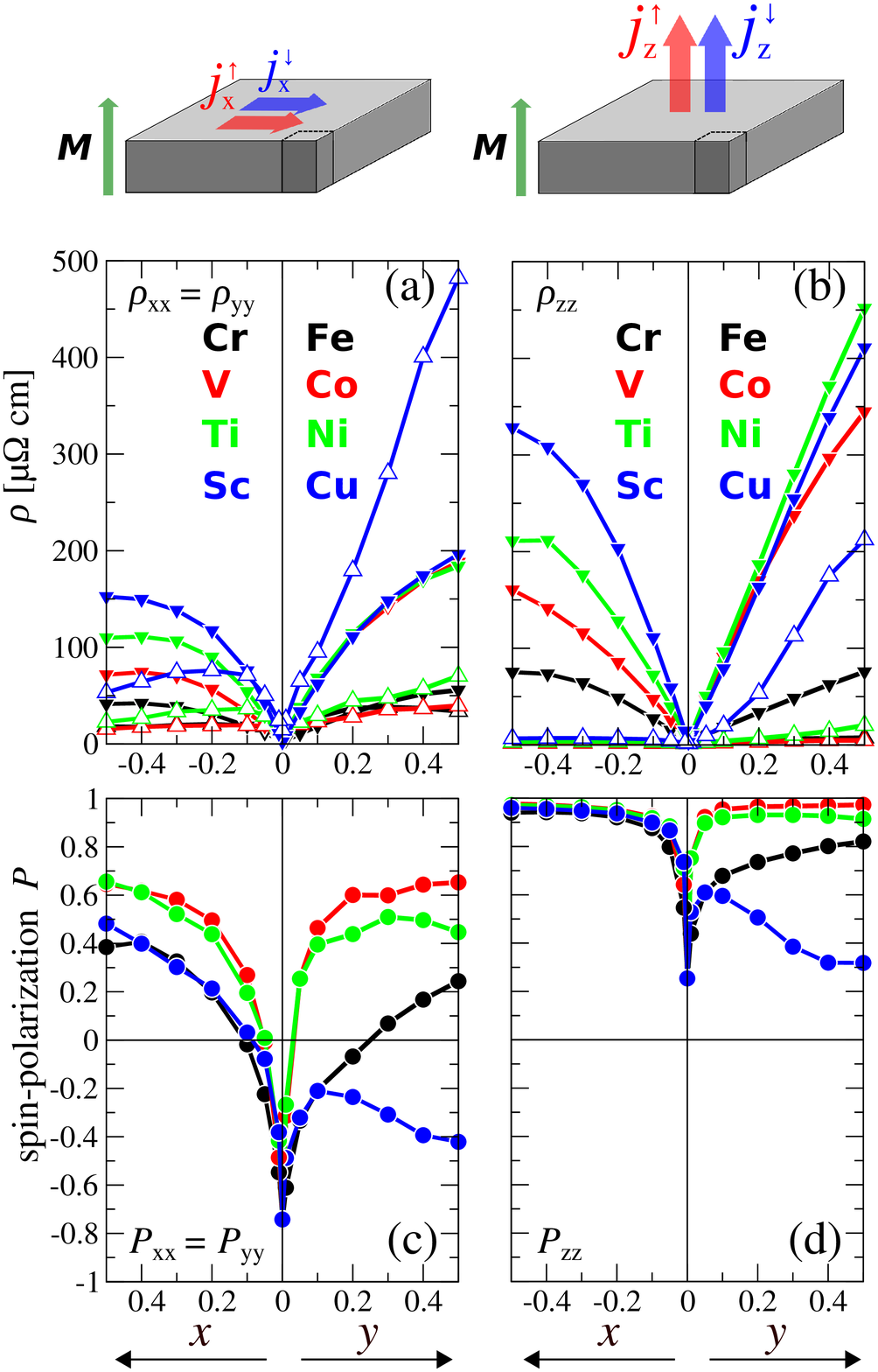}
\caption{\label{FIG:RHO-SP} Spatial components of residual resistivities,
  ${\rho_{\rm xx}=\rho_{\rm yy}}$~(a) and $\rho_{\rm zz}$~(b)  (their
  spin-projections, $\rho^{\uparrow(\downarrow)}$, are distinguished by
  the hollow up- and filled down-oriented triangles), computed as functions of 
  Mn$_{1-y}Y_y$ ($Y$ = Fe, Co, Ni, Cu) and Mn$_{1-x}X_x$ ($X$ = Cr, V, Ti,
  Sc) random substitutions; the substitution rates $x$ and $y$ increase in
  the directions indicated by the arrows. 
 Spatial components of spin-polarization, ${P_{\rm xx}=P_{\rm yy}}$~(c)
 and $P_{\rm zz}$~(d), are  derived from the spin-projections of the
 corresponding residual resistivity spatial components. The
 corresponding schemes  with  electric
 current, $\vec j$, magnetization, $\vec M$,  and the unit cell
 relative  orientations are shown above.}
\end{figure}

Since we examine the  tetragonal systems, we will
distinguish their properties along the in-plane (xy, or the  $ab$-plane of
the tetragonal lattice) and out-of-plane (along z, or the $c$-axis of the
tetragonal lattice) directions. Figures~\ref{FIG:RHO-SP}\,a 
and \ref{FIG:RHO-SP}\,c represent the in-plane transport properties (spin-projected resistivities and spin-polarizations, respectively),  whereas
Figures~\ref{FIG:RHO-SP}\,3b and 3d  represent the out-of-plane characteristics. In
all cases, we assume that $M$ is oriented along the z-axis because
of the  magnetocrystalline anisotropy. Here, we consider the diagonal
elements of the resistivity tensor, ${\rho_{\rm xx}=\rho_{\rm
    yy}\neq\rho_{\rm zz}}$, as being responsible for the direct current, ${\vec j}$
 (${j_\alpha=E_\alpha/\rho_{\alpha\alpha}}$, $\vec E$ is the external electric
field and  $\alpha$ is the spatial index $\rm x, y, or\ z$). The corresponding spatial components of spin-polarization are
defined as ${P_{\alpha\alpha}=\frac{\rho^{\downarrow}_{\alpha\alpha}-\rho^{\uparrow}_{\alpha\alpha}}{\rho_{\alpha\alpha}^{\downarrow}+\rho_{\alpha\alpha}^{\uparrow}}}$, where
$\rho_{\alpha\alpha}^{\uparrow,\,\downarrow}$ are the corresponding spin-projections. 

As follows from Figs.~\ref{FIG:RHO-SP}\,a and \ref{FIG:RHO-SP}\,b, 
with increasing chemical disorder induced by Mn-$X$ or Mn-$Y$ substitution,
almost all spatial/spin resistivity components grow monotonously within
${0\le x,y\le0.5}$ (with very few exceptions, e.g., for $\rho_{\rm
  xx}^{\uparrow}$ in the case of $X$ = Sc and Ti at higher $x$ rates, due to a
certain increase in  $n^{\uparrow}_{\rm F}$). This growth is most
efficient for small $x$ or $y$, and tends to saturate close to
${x\approx y\approx 0.5}$ (maximal disorder).
For the ordered Mn$_3$Ga, all resistivity components are exactly zero; for this reason, we estimate
its spin-polarization by extrapolating corresponding expressions to
${x,y\rightarrow0}$.  Here, we find that the pure Mn$_3$Ga represents a rather large
spin-polarization spatial anisotropy: ${P_{\rm xx}=P_{\rm  yy}\approx-0.75}$ (Fig.~\ref{FIG:RHO-SP}\,c), whereas ${P_{\rm zz}\approx+0.25}$  (Fig.~\ref{FIG:RHO-SP}\,d), which indicates that
the spin-polarization estimates for anisotropic systems (e.g.,~\cite{WBF+08}) based on a spin-resolved DOS
at $E_{\rm F}$  can be improved by considering the
spatially-resolved DOS in momentum space.
The resistivity spin component trends as functions of $x$ and $y$ show
that $\rho^{\downarrow}$  grows faster than $\rho^{\uparrow}$, which
essentially justifies the proposed scheme (see Sec.~\ref{sec:Random_fluctuations_design}). Indeed, both  $P_{\rm xx}$ and $P_{\rm
  zz}$, as  functions of $x$ or $y$, evolve towards larger positive
values. The absolute disorder-induced change of the $P_{\rm xx}$
component is very large, from $-0.75$ to approximately $+0.65$ (for
Mn-Ti, Mn-V, and Mn-Co substitutions), i.e. ${\Delta P_{\rm
    xx}\approx1.4}$. Despite the fact that this particular effect is not especially
interesting, since absolute spin-polarization does not increase,
it clearly demonstrates the importance of disorder. 

An interesting point worth mentioning is  that the disorder influence is stronger for those alloys in which the substituting type ($X$ or $Y$)
is further from Mn (e.g., in terms of the group or valence electrons
number), i.e., a larger potential difference leads to a stronger random
fluctuation amplitude. Indeed, for the types ``closest'' to Mn,
i.e., $X$ = Cr and $Y$ = Fe, the corresponding resistivity components are
comparably small (e.g., at ${x=y=0.5}$):
${\rho^{\uparrow}_{\rm xx}:18\sim33}$, 
${\rho^{\downarrow}_{\rm xx}:41\sim55}$,
${\rho^{\uparrow}_{\rm zz}:2.3\sim7.5}$, and
${\rho^{\downarrow}_{\rm zz}:75.3\sim75.7}$~$\mu\Omega$cm. At the same
time, for the types ``most distinct'' from Mn, i.e., $X$ = Sc and $Y$ = Cu,
certain resistivity components are much larger, but not always
comparable: ${\rho^{\uparrow}_{\rm xx}:53\nsim481}$, 
${\rho^{\downarrow}_{\rm xx}:152\sim197}$,
${\rho^{\uparrow}_{\rm zz}:6.9\nsim212}$, and
${\rho^{\uparrow}_{\rm zz}:327\sim411}$~$\mu\Omega$cm. This simply
indicates that, whereas substitution of Mn with ``too similar'' elements is not
yet appropriately efficient, substitution with ``too distinct'' elements
rapidly escapes  control, since the band structure is strongly affected not only in the sense of the Bloch-wave broadening, but also
in the sense of dispersion,  $E_{\rm F}$ position, etc. The most
inefficient situation is observed for Mn-Cu substitution. Here, $\rho^{\uparrow}$
and $\rho^{\downarrow}$ grow rapidly by achieving large absolute values, but their
ratios, and, thus, the spin-polarization, remain unsatisfactorily low. 
In contrast, substitution with ``intermediate'' elements, $X$ = V, Ti,
or $Y$ = Co, Ni, appears to be very efficient. Specifically, at ${x,y\approx0.1}$  the
out-of-plane spin-polarization achieves ${P_{\rm zz}\approx0.91-0.95}$ and grows further with increased disorder
rate. As follows from Fig.~\ref{FIG:RHO-SP}\,b, such growth is mainly due to the increase of
the $\rho_{\rm zz}^{\downarrow}$ component, whereas $\rho_{\rm
  zz}^{\uparrow}$ remains almost unaffected, as was supposed in
Sec.~\ref{sec:Random_fluctuations_design} during the discussion of the constructive 
disorder design. 

\begin{figure}
\centering
\includegraphics[clip,width=1.0\linewidth]{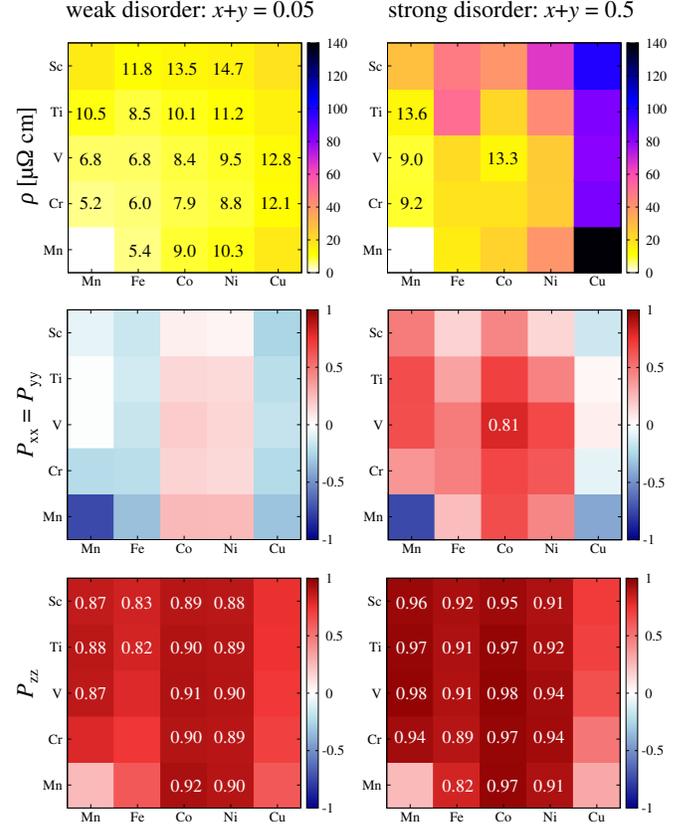}
\caption{\label{FIG:MAPS} Residual resistivities, $\rho$, (averaged
  over spins and spatial directions) and spin-polarization
  spatial components (${P_{\rm xx}=P_{\rm yy}}$, within xy-plane; $P_{\rm zz}$, out-of-plane) calculated for
  Mn$_{3-x-y}X_xY_y$ compositions ($X$ = Cr, V, Ti, Sc; $Y$ = Fe, Co, Ni, Cu; formally, we also include Mn). The total amount of substituted Mn is fixed
  to either ${x+y=0.05}$ (weak disorder, left-side) or
  to ${x+y=0.5}$ (strong disorder, right-side). Thus, for the
  strong disorder, if, for example, ${x=0}$, then ${y=0.5}$, and vice
  versa, or if both ${x,y>0}$,  then ${x=y=0.25}$. For the
  compositions with ${P>0.8}$ and ${\rho<15}$~$\mu\Omega$cm, the
  corresponding values are shown explicitly.}
\end{figure}

In order to also study the combined effects (in Mn$_{3-x-y}X_xY_y$ compositions), we plot the
computed transport characteristics,  ${P_{\rm xx}=P_{\rm yy}}$ and $P_{\rm zz}$, together
with the spatially- and spin-averaged (effective) resistivity,
${\rho=(2\rho_{\rm xx}+\rho_{\rm zz})/3}$, where
${\rho_{\alpha\alpha}=1/\left(1/\rho^{\uparrow}_{\alpha\alpha}+1/\rho^{\,\downarrow}_{\alpha\alpha}\right)}$ is
computed for the weak (${x+y=0.05}$) and strong (${x+y=0.5}$) disorder
regimes, as shown in Fig.~\ref{FIG:MAPS}.
It can be seen that, by moving towards  Sc and Cu, the effective resistivity tends
to increase. In both disorder regimes
this larger resistivity is exhibited by all Cu-containing 
 compositions, with a maximum of approximately
140~$\mu\Omega$cm for the Mn$_{2.5}$Cu$_{0.5}$Ga alloy. At the same time, the
spin-polarization of metals with large resistivity is always low,
since the spin component with higher resistivity cannot go far beyond
the Mooij limit, whereas the other spin component with lower resistivity is
already sufficiently high enough. On the other hand, alloys with lower
effective resistivities, such as Co- or V-containing compositions,  exhibit much
higher spin-polarization. Interestingly,  they show a noticeable
``complementary'' effect, seen for example  in the strongly disordered regime;  whereas for both
Mn$_{2.5}$Co$_{0.5}$Ga and Mn$_{2.5}$V$_{0.5}$Ga,  ${P_{\rm xx}\approx0.65}$, for the combined composition,
Mn$_{2.5}$V$_{0.25}$Co$_{0.25}$Ga, it is already 0.81. 
It is also instructive to admit the efficiency of the constructive disorder;
by moving from Mn$_3$Ga through the weakly disordered
Mn$_{2.95}$V$_{0.025}$Co$_{0.025}$Ga to
Mn$_{2.5}$V$_{0.25}$Co$_{0.25}$Ga, which is the ten times more strongly
disordered, $P_{\rm zz}$ evolves from 0.25 through 0.91 to 0.98,
respectively. This  means that, in order to achieve high spin-polarization,  the small
substitution rate is already sufficient.

\section{Summary and outlook}

As we have seen, an increase in spin-polarization is observed
almost for any type of Mn-TM substitution within Mn$_3$Ga, beginning
with Sc and ending with Ni. The exception is the Mn-Cu case, which leads to a
very strong random potential fluctuation affecting both spin
channels.  The important feature of Mn$_3$Ga is its tetragonal structure, which
causes a large anisotropy in its transport characteristics. As a
result, the spin-polarization along the in-plane and out-of-plane
directions evolves differently. Whereas the
disorder-induced change is strong in both directions, it is the most
constructive in the out-of-plane direction only.
At the same time, the constructive effect is achieved  immediately, by beginning with a
small Mn-TM substitution rate. This is suitable for spintronics elements 
exploiting magneto-resistance effects, as the unnecessarily high
resistance of the electrodes produces unwanted energy losses. Another
positive aspect of the presented scheme is the improvement in
spin-polarization specifically for the
``current-perpendicular-to-plane'' (CPP) setup. This improvement in
combination with perpendicular magnetic anisotropy  is applicable to many
state-of-the-art industrial developments. On the other hand, this
also means that the direct experimental proof of the proposed scheme,
at least on a Mn$_3$Ga basis, requires preparation of the single-crystalline structures
with effective stoichiometric control, which is rather sophisticated. In
this respect, an interesting future focus is constructive disorder design in
 cubic systems, as  high spin-polarization can
be expected in the isotropic case even for polycrystalline materials.
As we have seen, the basic feature which
provides the necessary conditions for such  disorder engineering,
is the presence of two antiparallel magnetic sublattices. Hence, suitable
cubic candidates for this research can be found within the same Mn-rich Heusler group.

\section{Appendix}
\label{sec:appendix}

All the present computations were performed using the fully relativistic
SPR-KKR (spin-polarized relativistic Korringa-Kohn-Rostoker) Green's
function method, using the generalized gradient approximation (GGA) in a
form proposed by Perdew,  Burke and Ernzerhof (PBE)~\cite{PBE96}. The chemical disorder was treated within the CPA~\cite{Sov67,But85}, as implemented in SPR-KKR. The electronic
structure is represented via the Bloch spectral function
(BSF), defined as a Fourier transform of the real-space Green's
function $G(\vec r,\vec r\,',E)$ with 
\begin{eqnarray*}
A(\vec k,E)&=&-\frac{1}{\pi N}~{\rm Im}\!\!\sum_{n,m=1}^N\!\!e^{i\vec k(\vec R_n - \vec
    R_m)}\\ &\times&\int d^3r \left\langle G(\vec r+\vec R_n,\vec r+\vec R_m, E)\right\rangle\,,
\end{eqnarray*}
where $\left\langle{}\right\rangle$ is the CPA average and  $\vec
R_{n,m}$ are the atomic site coordinates. $A(\vec k,E)$
can be interpreted as a $\vec k$-resolved DOS function, since
\begin{eqnarray*}
n(E) = \frac{1}{\Omega_{\rm BZ}}\int\limits_{\Omega_{\rm BZ}}d^3k\,A(\vec k,E)\,,
\end{eqnarray*}
with $n(E)$ indicating the  total DOS function, and $\Omega_{\rm BZ}$ the Brillouin zone volume.

\begin{acknowledgments}
Financial support from the Deutsche Forschungsgemeinschaft
(DfG) (research unit FOR~No.~1464 ``$\text{ASPIMATT}$'', project P~1.2-A) and
European Research Council Advanced Grant (ERC-AG) No.~291472 "IDEA Heusler'' is  gratefully acknowledged.
\end{acknowledgments}


\end{document}